\documentclass[10pt,preprintnumbers,showpacs]{revtex4}
\usepackage{graphicx,epsfig,axodraw}

\def\xslash#1{{\rlap{$#1$}/}}
\def\half{\frac{1}{2}}
\def\beq{\begin{equation}}
\def\eeq{\end{equation}}
\def\beqa{\begin{eqnarray}}
\def\eeqa{\end{eqnarray}}
\def\iar{\begin{array}{l}}
\def\ear{\end{array}}

\begin{document}

\title{Boson's field renormalization prescription}
\author{Yong Zhou}
\affiliation{Institute of High Energy Physics, Academia Sinica, P.O. Box 918(4), Beijing 100049, China, Email: zhouy@ihep.ac.cn}

\begin{abstract}
We discuss the problem of the present boson's field renormalization prescription induced by the imaginary parts of the unstable boson's propagation amplitudes and how to resolve it. 
\end{abstract}

\maketitle

\section{Introduction}

The field renormalization prescription has been present for a long time, but at present it encounters some new problems for unstable fermions \cite{c1,c2,c3}. The conventional field renormalization prescription isn't suitable for unstable fermions, but the revised field renormalization prescription in Ref.\cite{c4} leads to the physical amplitude gauge-parameter dependent \cite{c2}. Furthermore this prescription leads to the decay width of a physical process gauge-parameter dependent (see the appendix) which further proves the fermion field renormalization prescription of Ref.\cite{c4} isn't suitable for the standard model. D. Espriu et al. suggest to introduce two set independent Field Renormalization Constants (FRC) for the external-line fermion fields which can guarantee the physical amplitude gauge independent \cite{c2}. 

But for unstable boson the corresponding discussion is still not carried out. In this manuscript we will discuss this problem. In the follows we firstly discuss what problem exists in the present boson's field renormalization prescription. Then we discuss how to construct a reasonable boson's field renormalization prescription. In section 4 we illustrate the consistence of the present boson's field renormalization prescription with the gauge theory in standard model by the calculations of the physical amplitude of $Z\rightarrow d_i\bar{d}_i$, i.e. the gauge boson $Z$ decaying into a pair of down-type $i$ quarks. Lastly we give the conclusion.

\section{Problem of the present boson's field renormalization prescription}

Since the scalar boson's field renormalization prescription can be easily obtained from the vector boson's one, here we mainly discuss the vector boson's field renormalization prescription. The vector boson's FRC can be introduced as 
\beq
  \Phi_{0i}^{\mu}\,=\,\sum_j Z_{ij}^{\half}\Phi_j^{\mu}\,, \hspace{10mm}
  \Phi_{0i}^{\mu\dagger}\,=\,\sum_j \Phi_j^{\mu\dagger}\bar{Z}_{ji}^{\half}\,,
\eeq
where two set vector boson's FRC have been introduced. Obviously they should satisfy the hermitian conjugation relationship
\beq
  \bar{Z}_{ij}^{\half}\,=\,Z_{ji}^{\half\ast}\,.
\eeq
The renormalized vector boson's two-point function can be written as
\beq
  \begin{picture}(72,26)
      \Photon(0,5)(24,5){2}{4}
      \GCirc(36,5){12}{0.5}
      \Photon(48,5)(72,5){2}{4}
      \Text(10,15)[]{$j,\mu$}
      \Text(8,-5)[]{$p$}
      \Text(67,15)[]{$i,\nu$}
  \end{picture}\,\equiv\,i\hat{\Gamma}_{ij}^{\mu\nu}(p)\,=\,-i g^{\mu\nu}\sum_{k,l}
  \bar{Z}_{ik}^{\half}\left[ (p^2-m_{0k}^2)\delta_{kl}+\Sigma^T_{kl}(p^2) \right]
  Z_{lj}^{\half}-i\,p^{\mu}p^{\nu}\Sigma^L_{ij}(p^2)\,,
\eeq
where the shaded circle is the sum of the 1PI diagrams, $m_{0k}$ is the bare mass of vector boson $k$, and $\Sigma^T_{kl}$ is the transverse vector boson's two-point function removed the external-line FRC. The conventional vector boson's field renormalization prescription is
\beqa
  &&\hat{\Gamma}_{ij}^{\mu\nu}(p)\,\varepsilon_{\nu}(p)|_{p^2=m_i^2}\,=\,0\,,
  \hspace{10mm}
  \hat{\Gamma}_{ij}^{\mu\nu}(p)\,\varepsilon_{\nu}(p)|_{p^2=m_j^2}\,=\,0\,, \nonumber \\
  &&\lim_{p^2\rightarrow m_i^2}\frac{1}{p^2-m_i^2}\hat{\Gamma}_{ii}^{\mu\nu}(p)\,
  \varepsilon_{\nu}(p)\,=\,-\,\varepsilon^{\mu}(p)\,.
\eeqa
But it's well known that Eqs.(4) cannot be satisfied for unstable vector bosons since there is imaginary part in the transverse two-point function. In order to be suitable for unstable vector bosons and satisfy the constraint of Eq.(2) A. Denner revised Eqs.(4) as follows \cite{c4}
\beqa
  &&Re\,\hat{\Gamma}_{ij}^{\mu\nu}(p)\,\varepsilon_{\nu}(p)|_{p^2=m_i^2}\,=\,0\,,
  \hspace{10mm}
  Re\,\hat{\Gamma}_{ij}^{\mu\nu}(p)\,\varepsilon_{\nu}(p)|_{p^2=m_j^2}\,=\,0\,,
  \nonumber \\
  &&\lim_{p^2\rightarrow m_i^2}\frac{1}{p^2-m_i^2}Re\,\hat{\Gamma}_{ii}^{\mu\nu}(p)\,
  \varepsilon_{\nu}(p)\,=\,-\,\varepsilon^{\mu}(p)\,,
\eeqa
where $Re$ takes the real part of the two-point function. 

Of course Eqs.(5) are suitable for unstable vector bosons, and also satisfy the constraint of Eq.(2) \cite{c4}. But we find it leads to physical amplitudes gauge-parameter dependent. We calculate a physical process of $Z\rightarrow d_i\bar{d}_i$, i.e. the gauge boson $Z$ decaying into a pair of down-type $i$ quarks, to illustrate this problem. At one-loop level we have
\beqa
  {\cal M}(Z\rightarrow d_i\bar{d}_i)&=&[ -\frac{e}{6}\delta Z_{\gamma Z}+
  \frac{e(2 c_W^2+1)}{12 c_W s_W}(\delta Z_{ZZ}+\delta\bar{Z}^{dL}_{ii}+
  \delta Z^{dL}_{ii})]A_L-[\frac{e}{6}\delta Z_{\gamma Z}+\frac{e\,s_W}{6 c_W}
  (\delta Z_{ZZ}+\delta\bar{Z}^{dR}_{ii}+\delta Z^{dR}_{ii})]A_R \nonumber \\
  &+&{\cal M}^{amp}(Z\rightarrow d_i\bar{d}_i)\,,
\eeqa
where the vector boson's FRC have been expanded as $Z^{\half}_{ij}=\delta_{ij}+\half\delta Z_{ij}$, $\delta\bar{Z}^{dL}_{ii}$ et al. are $d_i$ quark's FRC \cite{c2}, $e$ is the electron's charge, $s_W$ and $c_W$ are the sine and cosine of the weak mixing angle, and
\beq
  A_L\,=\,\bar{u}(p_{d_i}){\xslash \epsilon}\gamma_L\nu(p_{\bar{d}_i})\,, \hspace{10mm}
  A_R\,=\,\bar{u}(p_{d_i}){\xslash \epsilon}\gamma_R\nu(p_{\bar{d}_i})\,,
\eeq
with $\gamma_L$ and $\gamma_R$ the left- and right- handed helicity operators, and ${\cal M}^{amp}$ is the amplitude of the amputated diagrams shown in Fig.1.
\begin{figure}[htbp]
\begin{center}
  \epsfig{file=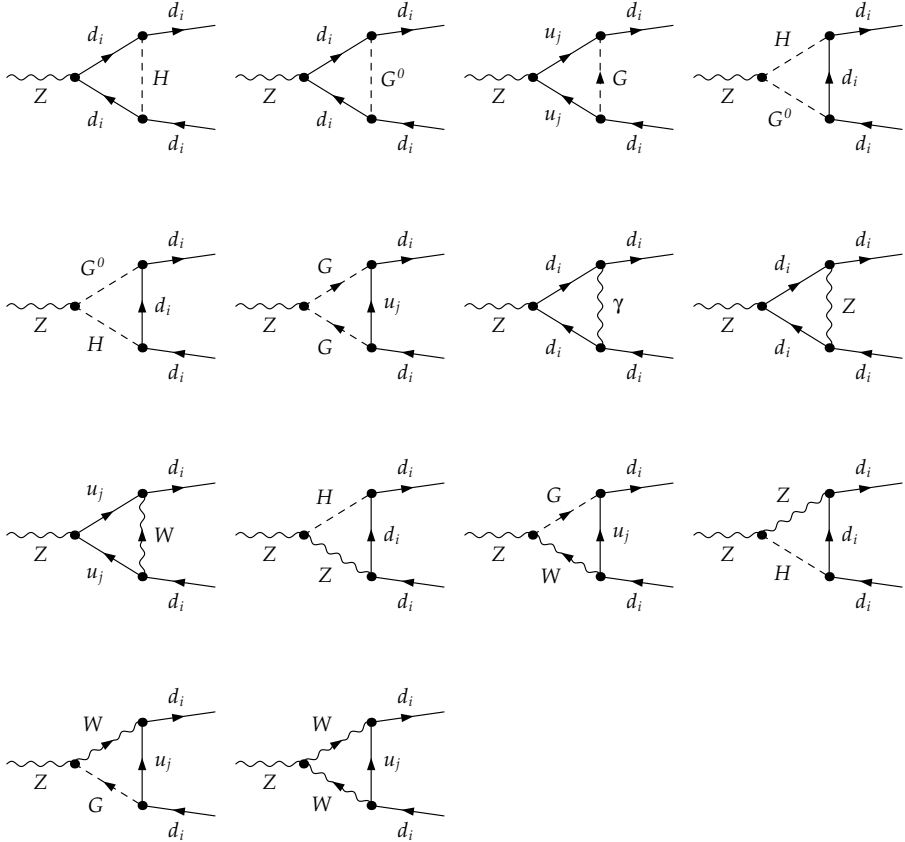,width=12cm} \\
  \caption{One-loop amputated diagrams of $Z\rightarrow d_i\bar{d}_i$.}
\end{center}
\end{figure}
Our numerical calculation has shown that the real part of the physical amplitude is gauge-parameter independent. So in the follows we only need to discuss the gauge dependence of the imaginary part of the physical amplitude. Firstly the gauge-dependent imaginary parts of $d_i$ quark's FRC are \cite{c2}
\beqa
  &&Im(\delta\bar{Z}^{dR}_{ii}+\delta Z^{dR}_{ii})|_{\xi}\,=\,0\,, \nonumber \\
  &&Im(\delta\bar{Z}^{dL}_{ii}+\delta Z^{dL}_{ii})|_{\xi}\,=\,
  \frac{e^2}{16\pi\,s_W^2\,x_{d,i}}\sum_j |V_{ji}|^2(x_{d,i}-x_{u,j}-\xi_W)\,B\,
  \theta[m_{d,i}-m_{u,j}-\sqrt{\xi_W}M_W]\,,
\eeqa
where the subscript $\xi$ takes the gauge-dependent part, $\theta$ is the Heaviside function, $m_{d,i}$ and $m_{u,j}$ are the masses of $d_i$ quark and up-type $j$ quark, $M_W$ and $\xi_W$ are the mass and gauge parameter of gauge boson $W$, $V_{ji}$ is the CKM matrix element \cite{c5}, $x_{d,i}=m_{d,i}^2/M_W^2$ and $x_{u,j}=m_{u,j}^2/M_W^2$, and
\beq
  B\,=\,\sqrt{\xi_W^2-2(x_{d,i}+x_{u,j})\xi_W+(x_{d,i}-x_{u,j})^2}\,.
\eeq
On the other hand using the {\em cutting rules} \cite{c6} we obtain
\beqa
  Im{\cal M}^{amp}(Z\rightarrow d_i\bar{d}_i)|_{\xi}&=&\left[
  \frac{e^3}{384\pi\,c_W^3 s_W^3}(1-4 c_W^2\xi_W)^{3/2}\theta[M_Z-2\sqrt{\xi_W}M_W]
  \right. \nonumber \\
  &-&\frac{e^3(2 c_W^2+1)}{192\pi\,c_W s_W^3\,x_{d,i}}\sum_j |V_{ji}|^2
  (x_{d,i}-x_{u,j}-\xi_W)\,B\,\theta[m_{d,i}-m_{u,j}-\sqrt{\xi_W}M_W] \nonumber \\
  &-&\left. \frac{e^3}{192\pi\,c_W^3 s_W}
  \left( (\xi_W-1)^2 c_W^4-2(\xi_W-5)c_W^2+1 \right)
  \,C\,\theta[M_Z-M_W-\sqrt{\xi_W}M_W] \right]A_L,
\eeqa
with $M_Z=M_W/c_W$ the mass of gauge boson $Z$ and
\beq
  C\,=\,\sqrt{(\xi_W-1)^2 c_W^4-2(\xi_W+1)c_W^2+1}\,.
\eeq
We note that the result of Eq.(10) coincides with the results of the conventional loop momentum integral algorithm \cite{c6} and the causal perturbative theory \cite{c7}. According to Eqs.(5) the boson's FRC $\delta Z_{\gamma Z}$  and $\delta Z_{ZZ}$ don't contain imaginary part. So from Eqs.(6,8,10) we obtain 
\beqa
  Im{\cal M}(Z\rightarrow d_i\bar{d}_i)|_{\xi}&=&\left[
  \frac{e^3}{384\pi\,c_W^3 s_W^3}(1-4 c_W^2\xi_W)^{3/2}\theta[M_Z-2\sqrt{\xi_W}M_W]
  \right. \nonumber \\
  &-&\left. \frac{e^3}{192\pi\,c_W^3 s_W}
  \left( (\xi_W-1)^2 c_W^4-2(\xi_W-5)c_W^2+1 \right)
  \,C\,\theta[M_Z-M_W-\sqrt{\xi_W}M_W] \right]A_L\,.
\eeqa
This means the physical amplitude of $Z\rightarrow d_i\bar{d}_i$ is gauge dependent under the boson's field renormalization prescription of Ref.\cite{c4}.

\section{Construct a reasonable boson's field renormalization prescription}

From the above discussion we find in order to keep physical amplitudes gauge invariant the constrain of Eq.(2) must be discarded. In fact the hermitian conjugation relationship of Eq.(2) is broken by the imaginary parts of the unstable boson's propagation amplitudes \cite{c2,c8} (we can also see this point in the following discussions). So in the follows we will discard the constraint of Eq.(2) and treat $\bar{Z}^{\half}$ and $Z^{\half}$ as independent quantities.

We firstly discuss how to construct the off-diagonal vector boson's FRC. Consider the integrality of physical amplitudes the off-diagonal field renormalization conditions of Eqs.(4) should keep unchanged, i.e.
\beq
  \hat{\Gamma}_{ij}^{\mu\nu}(p)\,\varepsilon_{\nu}(p)|_{p^2=m_i^2}\,=\,0\,,
  \hspace{10mm}
  \hat{\Gamma}_{ij}^{\mu\nu}(p)\,\varepsilon_{\nu}(p)|_{p^2=m_j^2}\,=\,0\,,
  \hspace{10mm} for\,\,i\,\not=\,j\,.
\eeq
From Eq.(3) the solutions of Eqs.(13) can be written as
\beqa
  \bar{Z}^{\half}_{ij}&=&\frac{1}{(m_{0j}^2-m_i^2)Z^{\half}_{jj}}\left( 
  \sum_{k\not=j}\bar {Z}^{\half}_{ik}(m_i^2-m_{0k}^2)Z^{\half}_{kj}+\sum_{k,l}
  \bar{Z}^{\half}_{ik}\Sigma^T_{kl}(m_i^2)Z^{\half}_{lj} \right)\,, 
  \hspace{10mm}for\,\,i\,\not=\,j\,, \nonumber \\
  Z^{\half}_{ij}&=&\frac{1}{(m_{0i}^2-m_j^2)\bar{Z}^{\half}_{ii}}\left( 
  \sum_{k\not=i}\bar {Z}^{\half}_{ik}(m_j^2-m_{0k}^2)Z^{\half}_{kj}+\sum_{k,l}
  \bar{Z}^{\half}_{ik}\Sigma^T_{kl}(m_j^2)Z^{\half}_{lj} \right)\,, 
  \hspace{10mm}for\,\,i\,\not=\,j\,.
\eeqa
One can easily see that the loop levels of the vector boson's FRC of the r.h.s. of Eqs.(14) are less than the ones of the l.h.s. of Eqs.(14). So we can use Eqs.(14) to determine the vector boson's off-diagonal FRC order by order by recursion algorithm. At one-loop level Eqs.(14) become
\beq
  \bar {Z}_{ij(1)}^{\half}\,=\,\frac{\Sigma^T_{ij(1)}(m_i^2)}{m_j^2-m_i^2}\,,
  \hspace{10mm}
  Z_{ij(1)}^{\half}\,=\,\frac{\Sigma^T_{ij(1)}(m_j^2)}{m_i^2-m_j^2}\,, \hspace{10mm}
  for\,\,i\,\not=\,j\,,
\eeq
where the subscript $(1)$ represents the one-loop-level parts of the quantities. From Eqs.(15) one can easily see that the hermitian conjugation relationship of Eq.(2) is broken by the imaginary part of the unstable vector boson's two-point function $\Sigma^T_{ij(1)}$. 

For the diagonal vector boson's FRC we cannot use the renormalization conditions of Eqs.(4) since the diagonal unstable vector boson's self energy cannot be renormalized to zero at physical mass point. We should determine it by the LSZ {\em reduction formula} \cite{c9}. However, the LSZ {\em reduction formula} has only been proven for stable particles \cite{c9}. So we need to postulate a generalization of the LSZ {\em reduction formula} to unstable particles. Under the postulation the LSZ {\em reduction formula} also holds true for unstable particles \cite{c8}. When expanding the vector boson's propagation amplitude at physical mass pole we have from Eq.(3)
\beq
  \frac{-i\,g^{\mu\nu}}{\sum_{k,l}\bar{Z}_{ik}^{\half}[ (p^2-m_{0k}^2)\delta_{kl}+
  \Sigma^T_{kl}(p^2) ]Z_{li}^{\half}}\,\begin{array}{c} \vspace{-7mm} \\ p^2
  \rightarrow m_i^2 \\ \vspace{-6mm} \\ -\hspace{-2mm}-\hspace{-2mm}-\hspace{-2mm}
  -\hspace{-2mm}-\hspace{-2mm}-\hspace{-2.5mm}-\hspace{-3mm}\longrightarrow \ear\,
  \frac{-i\,g^{\mu\nu}}{(p^2-m_i^2)( \sum_k\bar{Z}^{\half}_{ik}Z^{\half}_{ki}
  +\sum_{k,l}\bar{Z}^{\half}_{ik}\Sigma^{T\prime}_{kl}(m_i^2)Z^{\half}_{li} )+
  i\epsilon}\,,
\eeq
where $\Sigma^{T\prime}_{kl}=\partial\Sigma^T_{kl}/\partial p^2$, and $\epsilon=\sum_{k,l}\bar{Z}_{ik}^{\half}[(m_i^2-m_{0k}^2)\delta_{kl}+\Sigma^T_{kl}(m_i^2)]Z_{li}^{\half}/i$ is a small quantity. From Eq.(16) the unit residue condition requires
\beq
  \bar{Z}^{\half}_{ii}Z^{\half}_{ii}\,=\,1-\sum_{k\not=i}\bar{Z}^{\half}_{ik}
  Z^{\half}_{ki}-\sum_{k,l}\bar{Z}^{\half}_{ik}\Sigma^{T\prime}_{kl}(m_i^2)
  Z^{\half}_{li}\,.
\eeq
To one-loop level Eq.(17) becomes
\beq
  \bar{Z}^{\half}_{ii}Z^{\half}_{ii}\,=\,1-\Sigma^{T\prime}_{ii(1)}(m_i^2)\,.
\eeq
Obviously there is some freedom left in the definition of the diagonal vector boson's FRC.

In order to completely determine the diagonal vector boson's FRC we need to carefully investigate the feature of the renormalized vector boson's two-point function $\hat{\Gamma}_{ij}^{\mu\nu}$. In fact $\hat{\Gamma}_{ij}^{\mu\nu}$ is symmetric about its indexes, i.e.
\beq
  \hat{\Gamma}_{ij}^{\mu\nu}(p)\,=\,\hat{\Gamma}_{ji}^{\mu\nu}(p)\,, \hspace{10mm}
  for\,\,i\not=j\,.
\eeq
For the problem concerned we only need to consider the transverse part of $\hat{\Gamma}_{ij}^{\mu\nu}$. Putting Eqs.(15) into Eq.(3) we find that Eq.(19) is automatically satisfied at one-loop level since $\Sigma^T_{ij(1)}(p^2)=\Sigma^T_{ji(1)}(p^2)$. At two-loop level the transverse part of $\hat{\Gamma}_{ij}^{\mu\nu}(p)$ becomes from Eq.(3)
\beqa
  \hat{\Gamma}_{ij(2)}^T(p)&=&(p^2-m_i^2)Z^{\half}_{ij(2)}+
  \bar{Z}^{\half}_{ij(2)}(p^2-m_j^2)
  -\delta m_{i(1)}^2 Z^{\half}_{ij(1)}-\bar{Z}^{\half}_{ij(1)}\delta m^2_{j(1)} 
  \nonumber \\
  &+&\sum_{k\not=i,j}\bar{Z}^{\half}_{ik(1)}(p^2-m_k^2)Z^{\half}_{kj(1)} 
  +\bar{Z}^{\half}_{ii(1)}(p^2-m_i^2)Z^{\half}_{ij(1)}
  +\bar{Z}^{\half}_{ij(1)}(p^2-m_j^2)Z^{\half}_{jj(1)}
  +\Sigma^T_{ij(2)}(p^2) \nonumber \\
  &+&(\bar{Z}^{\half}_{ii(1)}+Z^{\half}_{jj(1)})\Sigma^T_{ij(1)}(p^2)
  +\sum_{k\not=i}\bar{Z}^{\half}_{ik(1)}\Sigma^T_{kj(1)}(p^2)
  +\sum_{k\not=j}\Sigma^T_{ik(1)}(p^2)Z^{\half}_{kj(1)}\,, \hspace{10mm} for\,\,i\not=j\,,
\eeqa
where the superscript $T$ in the l.h.s. of the equation takes the transverse part of the two-point function. In order to make Eq.(20) satisfy Eq.(19) for arbitrary momentum $p$ the terms in Eq.(20) containing the two-point function $\Sigma^T_{ij}(p^2)$ must be symmetric about the indexes $i$ and $j$. This leads to (from Eqs.(15) and the fact that $\Sigma^T_{ij}(p^2)$ is symmetric about the indexes $i$ and $j$)
\beq
  \bar{Z}^{\half}_{ii(1)}\,=\,Z^{\half}_{ii(1)}\,, \hspace{10mm}
  \bar{Z}^{\half}_{jj(1)}\,=\,Z^{\half}_{jj(1)}\,. 
\eeq
From Eqs.(14,15,21) we easily obtain
\beq
  \bar{Z}^{\half}_{ij(2)}\,=\,Z^{\half}_{ji(2)}\,, \hspace{10mm} for\,\,i\not=j\,.
\eeq
Obviously Eq.(20) satisfies Eq.(19) under Eqs.(15,21,22). From Eqs.(15,21) we also find that 
\beq
  \bar{Z}^{\half}_{ij(0)}\,=\,Z^{\half}_{ji(0)}\,, \hspace{10mm}
  \bar{Z}^{\half}_{ij(1)}\,=\,Z^{\half}_{ji(1)}\,.
\eeq

Eqs.(23) manifests that the two vector boson's FRC matrices $\bar{Z}^{\half}$ and $Z^{\half}$ satisfy transposition relationship between each other to one-loop level. In fact we can prove this relationship to all loop levels using the condition of Eq.(19) by recursion algorithm. Basing on the results of Eqs.(23) we only need to prove the conclusion: if the transposition relationship between $\bar{Z}^{\half}$ and $Z^{\half}$ is true to $n$-loop level, it will be also true to $n+1$-loop level. Under the condition
\beq
  \bar{Z}^{\half}_{ij(m)}\,=\,Z^{\half}_{ji(m)}\,, \hspace{10mm} 
  for\,\,m\,=\,0,1,\cdot\cdot\cdot,n\,,
\eeq
we can easily have from Eqs.(14)
\beq
  \bar{Z}^{\half}_{ij(n+1)}\,=\,Z^{\half}_{ji(n+1)}\,, \hspace{10mm} for\,\,i\not=j\,,
\eeq
i.e. the off-diagonal $n+1$-loop-level vector boson's FRC also satisfy the transposition relationship. Considering the $n+2$-loop-level part of $\hat{\Gamma}_{ij}^T(p)$
\beq
  \hat{\Gamma}_{ij(n+2)}^T(p)\,=\,\sum_k\sum_{u,v}\bar{Z}_{ik(u)}^{\half}
  (p^2-m_{0k}^2)_{(v)}Z_{kj(n+2-u-v)}^{\half}+\sum_{k,l}\sum_{u,v}\bar{Z}_{ik(u)}^{\half}
  \Sigma^T_{kl(v)}(p^2)Z_{lj(n+2-u-v)}^{\half}\,,
\eeq
in order to make Eq.(19) true for arbitrary momentum $p$ the terms containing $\Sigma^T_{kl(v)}(p^2)$ in Eq.(26) must be symmetric about the indexes $i$ and $j$, i.e.
\beq
  \sum_{k,l}\sum_{u}\bar{Z}_{ik(u)}^{\half}\Sigma^T_{kl(v)}(p^2)Z_{lj(n+2-u-v)}^{\half}
  \,=\,\sum_{k,l}\sum_{u}\bar{Z}_{jk(u)}^{\half}\Sigma^T_{kl(v)}(p^2)
  Z_{li(n+2-u-v)}^{\half}\,, \hspace{6mm} for\,\,v=1,\cdot\cdot\cdot,n+2\,.
\eeq
From Eq.(24) we find that Eq.(27) is automatically satisfied for $v\geq 2$. For $v=1$ the l.h.s. of Eq.(27) becomes
\beqa
  \sum_{k,l}\sum_{u}\bar{Z}_{ik(u)}^{\half}\Sigma^T_{kl(1)}(p^2)Z_{lj(n+1-u)}^{\half}
  &=&(\bar{Z}^{\half}_{ii(n+1)}+Z^{\half}_{jj(n+1)})\Sigma^T_{ij(1)}(p^2)
  +\sum_{k\not=i}\bar{Z}^{\half}_{ik(n+1)}\Sigma^T_{kj(1)}(p^2) \nonumber \\
  &+&\sum_{k\not=j}\Sigma^T_{ik(1)}(p^2)Z^{\half}_{kj(n+1)}
  +\sum_{k,l}\sum_{u=1}^n\bar{Z}_{ik(u)}^{\half}\Sigma^T_{kl(1)}(p^2)
  Z_{lj(n+1-u)}^{\half}\,.
\eeqa
From Eqs.(24,25) we find that the last three terms of the r.h.s. of Eq.(28) are symmetric about the indexes $i$ and $j$. Therefore in order to make Eq.(28) satisfy Eq.(27) we only need and must need the conditions
\beq
  \bar{Z}^{\half}_{ii(n+1)}\,=\,Z^{\half}_{ii(n+1)}\,, \hspace{10mm}
  \bar{Z}^{\half}_{jj(n+1)}\,=\,Z^{\half}_{jj(n+1)}\,.
\eeq
Since $i$ and $j$ are arbitrary, we have proven that the diagonal $n+1$-loop-level vector boson's FRC also satisfy the transposition relationship. Thus we have proven the recursion condition mentioned above. Combined Eqs.(23) this means to all loop levels
\beq
  \bar{Z}^{\half}_{ij}\,=\,Z^{\half}_{ji}\,.
\eeq 
Especially we have \cite{c8}
\beq
  \bar{Z}^{\half}_{ii}\,=\,Z^{\half}_{ii}\,.
\eeq
Obviously Eq.(30) is consistent with Eqs.(14) and satisfies Eq.(19) (see Eq.(3)).

From Eqs.(17,31) we have
\beq
  \bar{Z}_{ii}\,=\,Z_{ii}\,=\,1-\sum_{k\not=i}\bar{Z}^{\half}_{ik}Z^{\half}_{ki}
  -\sum_{k,l}\bar{Z}^{\half}_{ik}\Sigma^{T\prime}_{kl}(m_i^2)Z^{\half}_{li}\,.
\eeq
Thus we have completely determined the vector boson's FRC.

\section{Gauge invariance of physical amplitudes under the present boson's field renormalization prescription}

In this section we give an example of the calculation of physical amplitude to see whether the present boson's field renormalization prescription keeps physical amplitude gauge invariant. We calculate the physical amplitude of $Z\rightarrow d_i\bar{d}_i$. Part of the result has been calculated in section 2. Here we only need to calculate $\delta Z_{\gamma Z}$ and $\delta Z_{ZZ}$. From Eqs.(15) and Eqs.(18,31) we have at one-loop level
\beq
  \delta Z_{\gamma Z}\,=\,-\frac{2\Sigma^T_{\gamma Z}(m_Z^2)}{m_Z^2}\,, \hspace{10mm}
  \delta Z_{ZZ}\,=\,-\frac{\partial}{\partial p^2}\Sigma^T_{ZZ}(m_Z^2)\,.
\eeq
In Fig.2 we show the one-loop gauge-parameter dependent $Z\rightarrow\gamma$ diagrams which are used to calculate the gauge-parameter-dependent imaginary part of $\delta Z_{\gamma Z}$.
\begin{figure}[htbp]
\begin{center}
  \epsfig{file=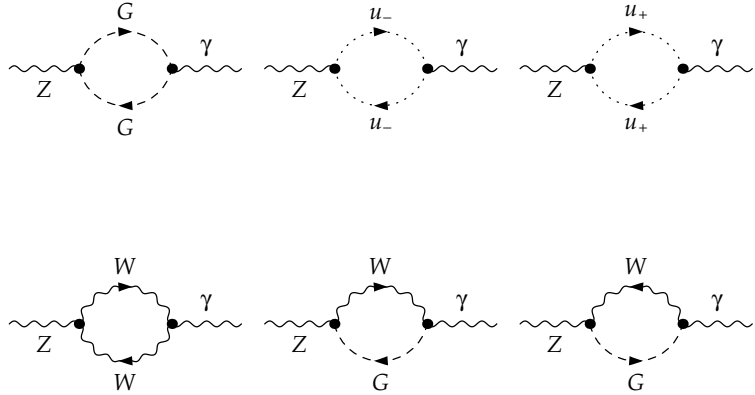,width=10cm} \\
  \caption{One-loop gauge-parameter dependent diagrams of $Z\rightarrow\gamma$
  which contain imaginary-part contribution.}
\end{center}
\end{figure}
Using the {\em cutting rules} we obtain from Eqs.(33)
\beqa
  Im\,\delta Z_{\gamma Z}|_{\xi}&=&\frac{e^2}{96\pi\,c_W^3 s_W}(1-4 c_W^2\xi_W)^{3/2}
  \theta[M_Z-2\sqrt{\xi_W}M_W] \nonumber \\
  &-&\frac{e^2\,s_W}{48\pi\,c_W^3}\left( (\xi_W-1)^2 c_W^4-2(\xi_W-5)c_W^2+1 \right)
  \,C\,\theta[M_Z-M_W-\sqrt{\xi_W}M_W]\,,
\eeqa
where $C$ has been shown in Eq.(11). On the other hand, in Fig.3 we show the one-loop gauge-parameter dependent $Z\rightarrow Z$ diagrams which are used to calculate the gauge-parameter-dependent imaginary part of $\delta Z_{ZZ}$.
\begin{figure}[htbp]
\begin{center}
  \epsfig{file=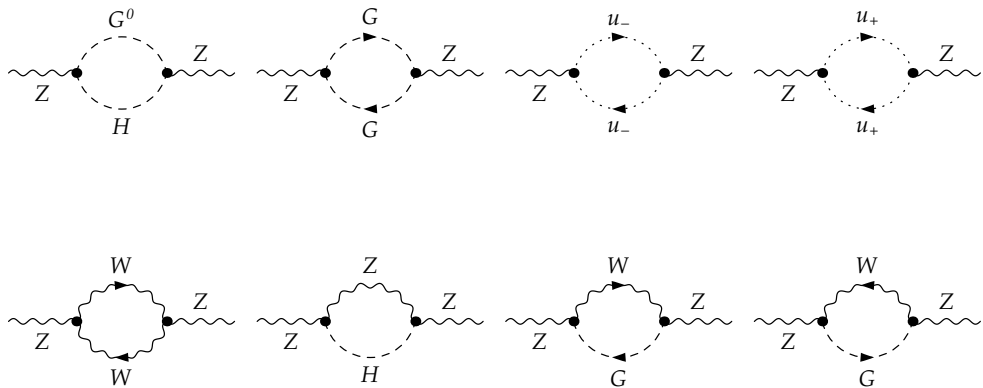,width=13cm} \\
  \caption{One-loop gauge-parameter dependent diagrams of $Z\rightarrow Z$
  which contain imaginary-part contribution.}
\end{center}
\end{figure}
Using the {\em cutting rules} we obtain from Eqs.(33)
\beqa
  Im\,\delta Z_{ZZ}|_{\xi}&=&-\frac{e^2}{96\pi\,c_W^2 s_W^2}(1-4 c_W^2\xi_W)^{3/2}
  \theta[M_Z-2\sqrt{\xi_W}M_W] \nonumber \\
  &+&\frac{e^2}{48\pi\,c_W^2}\left( (\xi_W-1)^2 c_W^4-2(\xi_W-5)c_W^2+1 \right)
  \,C\,\theta[M_Z-M_W-\sqrt{\xi_W}M_W]\,.
\eeqa 
Putting Eqs.(34,35) and Eqs.(8,10) into Eq.(6) we finally obtain
\beq
  Im{\cal M}(Z\rightarrow d_i\bar{d}_i)|_{\xi}\,=\,0\,.
\eeq
This means the present boson's field renormalization prescription keeps the physical amplitude of $Z\rightarrow d_i\bar{d}_i$ gauge-parameter independent.

\section{Conclusion}

In summary, we firstly discuss the present boson's field renormalization prescriptions and find the prescription of Ref.\cite{c4} leads to the physical amplitude gauge-parameter dependent. Then we postulate a generalization of the LSZ {\em reduction formula} to unstable particles and use the symmetry of the boson's two-point function about its particle's indexes to construct a reasonable boson's field renormalization prescription. The calculation of the physical amplitude of $Z\rightarrow d_i\bar{d}_i$ shows that the present boson's field renormalization prescription is consistent with the gauge theory in standard model.

\vspace{5mm} {\bf \Large Acknowledgments} \vspace{2mm} 

The author thanks Prof. Xiao-Yuan Li for his useful guidance and Prof. Cai-dian Lu for the fruitful discussions with him.

\vspace{5mm} {\bf \Large Appendix} \vspace{2mm}

In the appendix we calculate the gauge dependence of the decay width of $t\rightarrow c\,Z$, i.e. top quark decaying into charm quark and gauge boson Z, under the fermion field renormalization prescription of Ref.\cite{c4}. At one-loop level we have
\beq
  {\cal M}(t\rightarrow c\,Z)\,=\,\frac{e(4 s_W^2-3)}{12 s_W c_W}
  (\delta Z^L_{ct}+\delta\bar{Z}^L_{ct})\bar{c}\,{\xslash \epsilon^{\ast}}\gamma_L\,t+
  \frac{e\,s_W}{3 c_W}(\delta Z^R_{ct}+\delta\bar{Z}^R_{ct})\bar{c}\,{\xslash \epsilon^{\ast}}\gamma_R\,t+
  {\cal M}^{amp}(t\rightarrow c\,Z)\,,
\eeq
where ${\cal M}^{amp}$ is the amplitude of the one-loop amputated diagrams shown in Fig.4, and the quark's FRC $\delta Z^L_{ct}$ et al. are listed in Ref.\cite{c2} which satisfy the relationship $\delta\bar{Z}^L_{ij}=\delta Z^{L\dagger}_{ij}$ and $\delta\bar{Z}^R_{ij}=\delta Z^{R\dagger}_{ij}$ under the fermion field renormalization prescription of Ref.\cite{c4}.
\begin{figure}[htbp]
\begin{center}
  \epsfig{file=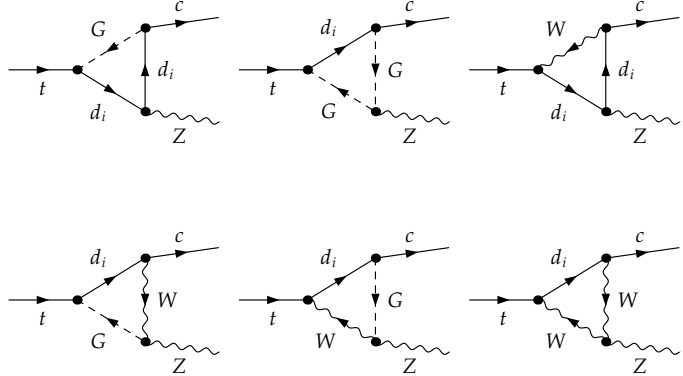,width=9cm} \\
  \caption{One-loop amputated diagrams of $t\rightarrow c\,Z$.}
\end{center}
\end{figure}
our numerical result has shown the {\em quasi-real part}, which takes the real part of the loop momentum integrals appearing in the amplitude but not of the coupling constants appearing there \cite{c4}, of Eq.(37) is gauge-parameter independent. So we only need to calculate the gauge dependence of the {\em quasi-imaginary part}, which takes the imaginary part of the loop momentum integrals appearing in the amplitude but not of the coupling constants appearing there \cite{c6}, of Eq.(37). According to Eqs.(3.20) of Ref.\cite{c4} the {\em quasi-imaginary parts} of the quark's FRC are equal to zero, so we only need to calculate the {\em quasi-imaginary part} of ${\cal M}^{amp}(t\rightarrow c\,Z)$. Using the {\em cutting rules} \cite{c6} we obtain 
\beqa
  \tilde{Im}{\cal M}(t\rightarrow c\,Z)|_{\xi}\,=\hspace{-3mm}&&\bar{c}\,
  {\xslash \epsilon^{\ast}}\gamma_L\,t\,\sum_i\frac{V_{2i}V^{\ast}_{3i}\,e^3
  (4 s_W^2-3)}{384\pi\,c_W\,s_W^3}\,\Bigl [ \nonumber \\
  &&\frac{x_c-\xi_W-x_{d,i}}{x_c}\sqrt{x_c^2-2(\xi_W+x_{d,i})x_c+(\xi_W-x_{d,i})^2}\,
  \theta[m_c-m_{d,i}-M_W\sqrt{\xi_W}] \nonumber \\
  +\hspace{-3mm}&&\left. \frac{x_t-\xi_W-x_{d,i}}{x_t}
  \sqrt{x_t^2-2(\xi_W+x_{d,i})x_t+(\xi_W-x_{d,i})^2}\,
  \theta[m_t-m_{d,i}-M_W\sqrt{\xi_W}] \right],
\eeqa
where $\tilde{Im}$ takes the {\em quasi-imaginary part} of the amplitude, $V_{2i}$ and $V_{3i}$ are the CKM matrix elements \cite{c5}, $m_c$ and $m_t$ are the masses of charm quark and top quark, and $x_c=m_c^2/M_W^2$, $x_t=m_t^2/M_W^2$. We note that the result of Eq.(38) coincides with the results of the conventional loop momentum integral algorithm \cite{c6} and the causal perturbative theory \cite{c7}. Since there is no tree level contribution, the result of Eq.(38) directly have contribution to the cross section of the physical process. In other words the decay width of $t\rightarrow c\,Z$ is gauge-parameter dependent under the fermion field renormalization prescription of Ref.\cite{c4}.

\end{document}